\documentclass[a4paper]{article}
\usepackage{environ}

\NewEnviron{FitToWidth}[1][\textwidth]{
\begin{center}
  \makebox[#1]{\resizebox{#1}{!}{\BODY}}
\end{center}
}
\usepackage{tikz}
\usepackage{xcolor}
\usepackage[colorlinks=true, allcolors=black]{hyperref}

\definecolor{CHN_CRY}{HTML}{437DA3}
\definecolor{CHN_FUS}{HTML}{00B1F7}
\definecolor{CHN_BAB}{HTML}{0019F7}
\definecolor{FAN_CDS}{HTML}{F70000}
\definecolor{FAN_FAN}{HTML}{C53BD1}
\definecolor{FAN_LAU}{HTML}{E0B8B8}
\definecolor{FAN_SNG}{HTML}{D67AA0}
\definecolor{MAN_CDS}{HTML}{3A6901}
\definecolor{MAN_MAN}{HTML}{1AFF00}
\definecolor{MAN_LAU}{HTML}{BBE0B8}
\definecolor{MAN_SNG}{HTML}{1ADB71}
\definecolor{CXN_CXN}{HTML}{F78C00}

\usepackage{float}
\usepackage{graphicx}
\usepackage{caption}
\usepackage{subcaption}
\usepackage{INTERSPEECH2022}

\title{Visualizations of Complex Sequences of Family-Infant Vocalizations Using Bag-of-Audio-Words Approach Based on Wav2vec 2.0 Features}
\name{Jialu Li$^{1,2}$, Mark Hasegawa-Johnson$^{1,2}$, Nancy L. McElwain$^{2,3}$}

\address{
  $^1$Department of Electrical and Computer Engineering, University of Illinois\\ 
  $^2$Beckman Institute for Advanced Science and Technology, University of Illinois \\
  $^3$Department of Human Development and Family Studies, University of Illinois}
\email{jialuli3@illinois.edu, jhasegaw@illinois.edu, mcelwn@illinois.edu}

\begin{document}

\maketitle
\begin{abstract}
In the U.S., approximately 15-17\% of children 2-8 years of age are estimated to have at least one diagnosed mental, behavioral or developmental disorder. However, such disorders often go undiagnosed, and the ability to evaluate and treat disorders in the first years of life is limited. To analyze infant developmental changes, previous studies have shown advanced ML models excel at classifying infant and/or parent vocalizations collected using cell phone, video, or audio-only recording device like LENA. In this study, we pilot test the audio component of a new infant wearable multi-modal device that we have developed called LittleBeats (LB). LB audio pipeline is advanced in that it provides reliable labels for both speaker diarization and vocalization classification tasks, compared with other platforms that only record audio and/or provide speaker diarization labels. We leverage wav2vec 2.0 to obtain superior and more nuanced results with the LB family audio stream.
We use a bag-of-audio-words method with wav2vec 2.0 features to create high-level visualizations to understand family-infant vocalization interactions. We demonstrate that our high-quality visualizations capture major types of family vocalization interactions, in categories indicative of mental, behavioral, and developmental health, for both labeled and unlabeled LB audio.

\end{abstract}

\section{Introduction}

Among U.S. children between 2 and 8 years of age, approximately 15-17\% are estimated to have at least one diagnosed mental, behavioral or developmental disorder
(MBDD)~\cite{bitsko2016health}, yet the ability to evaluate and treat MBDDs in the first years of life is limited. Child mental health problems, in part, emerge from daily interactions with family members that are repeated and reinforced over time. One well-known example is parent-child ``coercive cycles'' \cite{coercive}: when parents respond harshly to a child's negative affect or behavior, the child may respond with anger and resistance. Such parent-child interactions serve to amplify parent and child negative behaviors, which might reinforce the child's negative behaviors and lead to behavioral or emotional problems for the child in the long run. Sibling-infant interactions also play an important role in infant development~\cite{dunn1983sibling}. Although siblings often show positive behaviors with each other, conflicts commonly happen at home partly due to sibling's jealousy and rivalry for the loss of parents' attention~\cite{kolak2011sibling,volling2002emotion}. Thus, although the primary emphasis of empirical work has been on the mother-infant dyad, it is critical to consider the larger family context in understanding the development of behavioral or emotional disturbances during the first years of life.

Although previous psychological findings indicate attributes of family-child interaction that are correlated with later mental health problems among children~\cite{dadds1992childhood,ward2016parent}, it remains challenging to build sophisticated machine learning (ML) models to automatically predict and analyze critical home-life interaction patterns to better support child mental health outcomes. 
Typically, recordings of parent-child interactions are observed during brief tasks conducted in a controlled laboratory setting and, thus, contain limited examples of interaction patterns. In contrast, training robust ML models requires many labeled examples.  
From a clinical perspective, if such ML models existed, parents, physicians and early childhood educators would be better able to identify patterns of interaction that have been correlated, in previous scientific studies, with negative mental health outcomes. 

To analyze family interactions, in the past, researchers or parents have recorded family audio at home or laboratory using a cell phone, video camera, or an audio-only recording device like the Language Environment Analysis device (LENA) \cite{LENA}. 
Previous studies have demonstrated the ability of ML models to automatically classify infant and parent vocalizations collected from those audio-recording devices~\cite{anders2020automatic,anders2020comparison,ji2021review}. 
Because data annotation is a time-consuming and labor-intensive task, past studies have also experimented with transfer learning techniques by incorporating external and relevant datasets for additional training to further improve the classification performance~\cite{li2021analysis,gujral2019leveraging}. For example, wav2vec 2.0 (W2V2) \cite{wav2vec}, which uses unsupervised pre-training from $\sim$ 52k-hours of unlabeled raw wav audio, excels at multiple speech processing tasks, including speech-to-text~\cite{li2021accent,schneider2019wav2vec}, speech emotion recognition~\cite{pepino2021emotion,yuan2021role}, and speaker verification~\cite{wang2021fine}. 
The unsupervised W2V2 model has also been beneficially applied to several resource-poor down-stream speech processing tasks by fine-tuning on a small amount of target labeled data.


In this study, we pilot test the audio component of a new infant wearable multi-modal device we developed called Littlebeats$^{\text{TM}}$ (LB). LB audio pipeline is advanced in that it automatically provides reliable labels of speaker diarization and vocalization classifications for family members, including infant, parents, and siblings. In contrast, devices used in studying infant developmental changes in the past typically provide audio only or speaker diarization labels only with moderate reliability when compared with manually labeled recordings~\cite{bulgarelli2020look}.
To overcome the problem of scarcity of labeled LB audio data, we leverage W2V2 to automatically detect and classify infant and parent vocalizations in the LB audio stream.
The current ML community has limited studies analyzing family interactions, or understanding
complex event sequences in any other type of audio besides adult speech. To fill this research gap, we further analyze family vocalization interaction patterns by creating high-level visualizations of long sequences of family vocalizations based on extracted W2V2 features for both labeled and unlabeled daylong audio recordings in the home environment. 
This paper makes two contributions to science: (1) pilot testing our new wearable LB device with expanded functionalities, and (2) advancing ML audio understanding of sequences of family-infant vocalization interactions.

\section{Data}
\label{sec:data}

For this study, we recruited families from the community with study flyers distributed at child care centers, pediatric clinics, and local community organizations or online forums visited by families of infants (e.g., public libraries, parenting groups). Thirteen two-parent families with infants (8 males) between the ages of 1.1 months and 14 months participated. Nine infants were first-born, 2 infants were second-born, and 2 infants were third- or later-born. On average, parents were well-educated; 10 mothers and 4 fathers reporting having advanced degrees. 

All study procedures were approved by the Institutional Review Board at the University of Illinois at Urbana-Champaign.
Each infant wore the LB device during the day for two or three days. For the purpose of the manual annotation task, all LB recordings were separated into 10-mins segments. 
Because continuous annotation of the audio recordings is time- and labor-intensive, we were not able to complete annotations for all audio segments. Instead, human labelers annotated two or three 10-mins segments, selected according to the highest active vocalization rates computed by a statistical voice activity detector, for each family.
Human labelers manually labeled key child (CHN), female adult (FAN), male adult (MAN), and other child/sibling (CXN) vocalizations using Praat software~\cite{boersma2006praat}, with cross-labeler validation at a precision of 0.2 seconds. Ten percent of selected segments were double-coded, and the inter-rater reliability score (Cohen's kappa score) is 0.90 for CHN, 0.87 for FAN, 0.83 for MAN, and 0.79 for CXN. Child vocalizations were manually labeled as cry (\texttt{CRY}), fuss (\texttt{FUS}), and babble (\texttt{BAB}); adult vocalizations (MAN and FAN) were manually labeled as child-directed speech (\texttt{CDS}), adult-directed speech (\texttt{ADS}), laugh (\texttt{LAU}), and singing/rhythmic speech (\texttt{SNG}). The Cohen's kappas were 0.74 for CHN, 0.86 for FAN, and 0.74 for MAN. In total, we obtained 45 annotated 10-mins segments. The distribution of infant age for the 45 annotated segments are shown in Table \ref{tab:data_distribution}.

To prepare audio data for training W2V2 model, we labeled the audio stream in intervals of 2 seconds starting every 0.2 second. The label of each 2s interval is determined by the temporal majority of human annotations (e.g., if less than 50\% of the interval is coded as \texttt{BAB} and more than 50\% of interval is labeled as \texttt{CRY}, the interval is labeled as \texttt{CRY}). To reduce speaker diarization errors, intervals labeled with more than one speaker are discarded, and an interval is labeled as silence if the energy is below a certain threshold. 
The energy threshold is set just below the minimum energy of CHN segments in our corpus; thus CXN, FAN and MAN speech that happen far from CHN is excluded from the training corpus.
In this way, we intentionally select vocalizations that are close to the CHN for training the W2V2 model and ignore those that have lower energy or can be easily confused with background noise. All labeled data is randomly split into training, development, and testing sets in a ratio of 7:1:2.
\begin{table}
  \caption{Distribution of infant age for the annotated segments.}
  \label{tab:data_distribution}
  \centering
  \vspace{-0.2cm}
  \setlength{\tabcolsep}{5.0pt}
  \begin{tabular}{c|c|c}
    \toprule
    age (months) & \# participants & \# annotated recordings\\ 
    \midrule
    1.1-3.5 & 3 & 10\\
    3.5-6 & 3 & 10\\
    6-10 & 2 & 7\\
    10-14 & 5 & 18\\
    \bottomrule
  \end{tabular}
  \vspace{-0.4cm}
\end{table}

\section{Model \& Experimental Results}
We followed the typical fine-tuning procedure of training W2V2 model by simply feeding our labeled dataset to the neural-networks. We integrated 4 output tiers on top of W2V2 architecture, including a speaker diarization (SD) tier and 3 vocalization classification tiers. The SD tier learns to detect speaker as silent or one of CHN, FAN, MAN, and CXN; if not silent, the corresponding vocalization tier learns to classify the vocalization type. We implemented the W2V2 model using the SpeechBrain framework~\cite{speechbrain}, and our code is publicly available at Github\footnote[1]{\href{https://github.com/jialuli3/speechbrain/tree/infant-voc-classification/recipes/wav2vec\_kic}{https://github.com/jialuli3/speechbrain/tree/infant-voc-classification/recipes/wav2vec\_kic}}. W2V2 model accepts raw wav audio stream for 2s intervals starting every 0.2s, and the outputs of each interval are averaged along time dimension before feeding to four output tiers. The W2V2 system was fine-tuned using our labeled training set for 60 epochs until convergence. Adam optimizer with learning rates of output tiers and wav2vec model starting from 1e-4 and 1e-5 respectively was used; scheduler with new-bob technique was used to anneal learning rates based on development set performance after each epoch. The epoch with the best Cohen's kappa scores of SD tier on development set was used for final evaluation on testing set and extraction of W2V2 features for visualizations. Table \ref{tab:results} presents results on all four output tiers for different metrics. Kappa scores were 0.67 for the SD tier and 0.84 or greater for all three vocalization tiers, which suggests W2V2's potential robustness and strength in labeling family vocalizations in LB home recordings.

\begin{table}[H]
  \centering
  \vspace{-0.2cm}
  \setlength{\tabcolsep}{4.0pt}
   \caption{\small Accuracy, macro F1-scores, and Cohen's kappa scores for speaker diarization (SD) and three vocalization classification tiers (CHN, FAN, MAN) on testing set.}
    \label{tab:results}
      \vspace{-0.2cm}
  \begin{tabular}{l||ccc}
    \toprule
     tier & Acc & F1 & kappa \\
    \midrule\midrule
    SD & 0.838 & 0.725 & 0.67 \\
    \midrule 
    CHN & 0.898 & 0.888 & 0.84 \\
    FAN & 0.916 & 0.863 & 0.85\\
    MAN & 0.920 & 0.825 & 0.86\\
    \bottomrule
  \end{tabular}
\label{tab:classify}
  \vspace{-0.3cm}
\end{table} 

\section{Visualizations}
\subsection{Bag-of-audio-words (BoAW) approach for feature extraction over long periods of audio}
The Bag-of-words (BoW) model~\cite{zhang2010understanding} is commonly used in natural language processing and information retrieval tasks to extract simplified representations over texts. 
Similar BoW applications in audio domain (BoAW) are also well established. The key insight of BoAW algorithm is to learn codebooks (audio words) over all input features and then transform variable-length audio chunks into fixed-dimensional histograms (BoAW) of the learned codebooks. BoAW has been shown successful for training classifiers in acoustic event classification~\cite{zhang2018unsupervised}, multimedia event detection~\cite{pancoast2012bag}, and speech emotion recognition ~\cite{schmitt2016border,han2018bags}. 
In recent Interspeech Paralinguistics Challenges, BoAW and support vector machines with input of MFCCs and/or paralinguistics feature set, consisting of functionals (statistics) over low-level descriptor of acoustic features, are widely used as 
a baseline model for classifying various vocalizations, such as baby sounds~\cite{schuller2019interspeech,schuller2018interspeech}, snoring~\cite{schuller2017interspeech}, and COVID-19 cough~\cite{schuller2021interspeech}.

In this study, we modify typical BoAW workflow to efficiently extract W2V2 features spanning over a relatively long period of audio stream. Figure \ref{fig:workflow} shows the modified workflow. Given an audio stream, W2V2 features are extracted for intervals of 2 seconds starting every 0.2 second. Then all extracted W2V2 features are fed into k-means cluster algorithms to create codebooks/audio words (centroids of the k-means algorithms). Codebook size is set as 50 for efficient computation, as we find larger codebook sizes yield similar visualizations. 
After codebooks are generated, soft probability is used to assign each W2V2 vector to the 5 nearest codebooks based on Euclidean distances. For sequences of 30s W2V2 features, histograms of codebooks are generated and normalized by dividing each term frequency (TF) by the sum of all TFs. 
Those normalized histograms of codebooks are our BoAW. 
Principal Component Analysis (PCA) and T-distributed stochastic neighbor embedding (T-SNE)~\cite{van2008visualizing} are used to reduce 50-dimensional BoAW to 2-dimensional features for visualizations. We implement BoAW model using the OpenXBOW ~\cite{schmitt2017openxbow} toolkit and create visualizations using the scikit-learn package ~\cite{scikit-learn}. We use default setting of T-SNE in scikit-learn.
\begin{figure}
  \centering
  \includegraphics[width=0.75\linewidth]{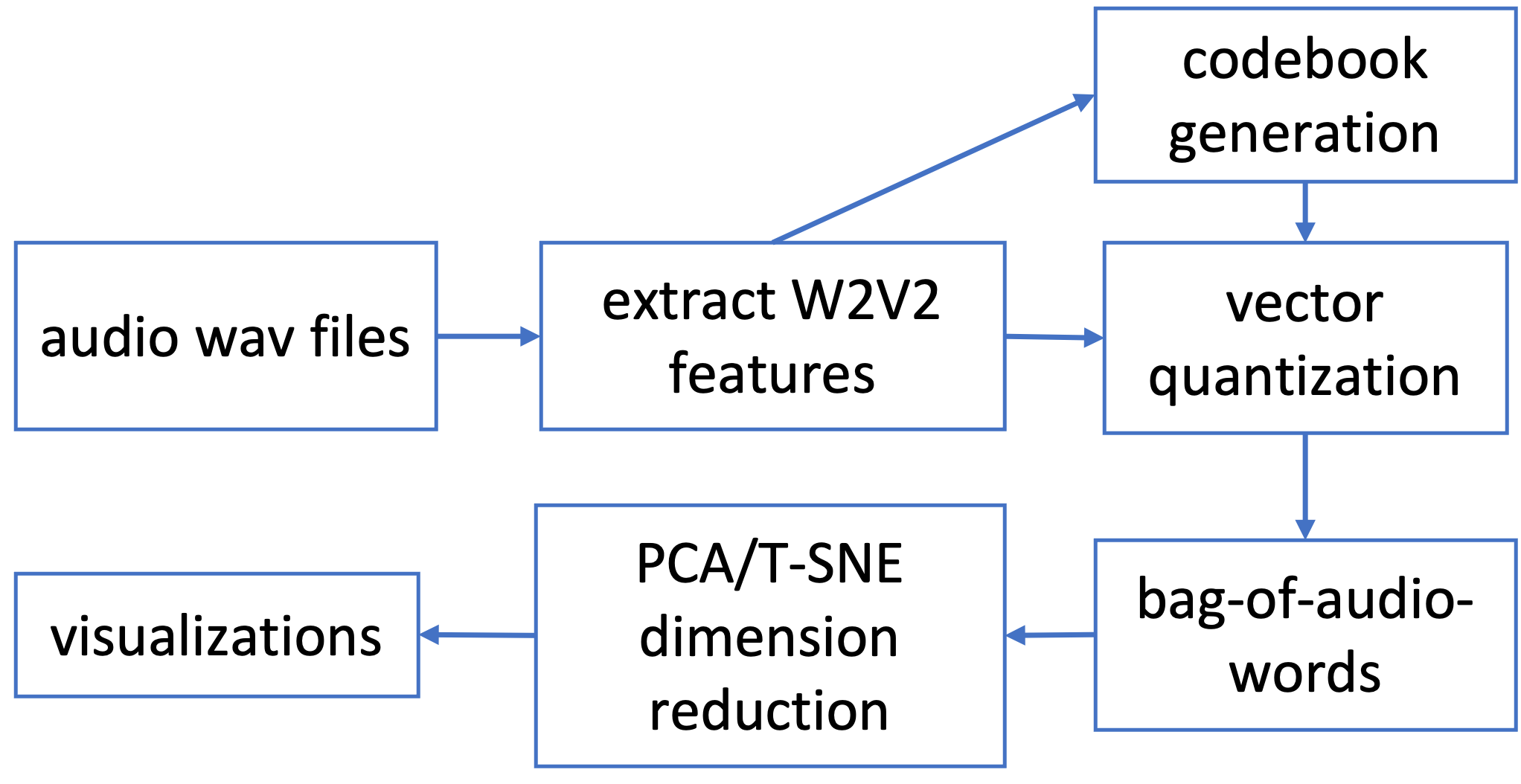}
    \vspace{-0.2cm}
  \caption{Algorithm workflow of BoAW approach. }
  \label{fig:workflow}
  \vspace{-0.65cm}
\end{figure}
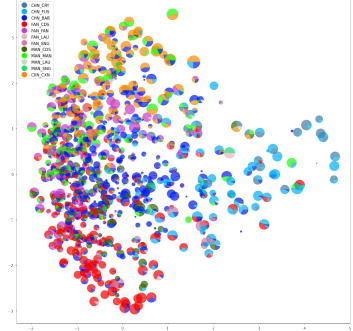
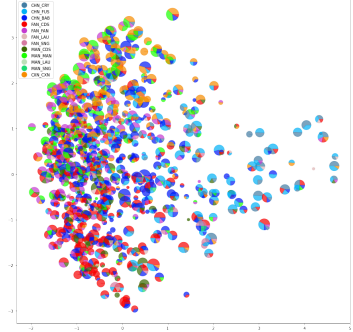
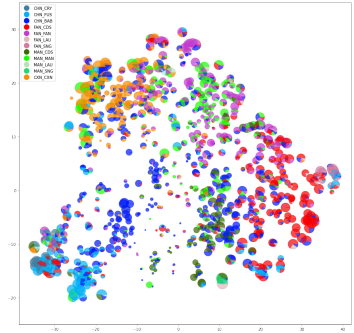
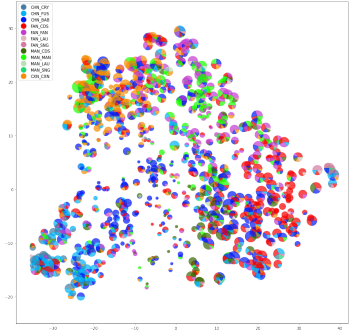
\begin{figure}
        \centering
        \begin{subfigure}[b]{0.27\textwidth}
          \includegraphics[width=\textwidth]{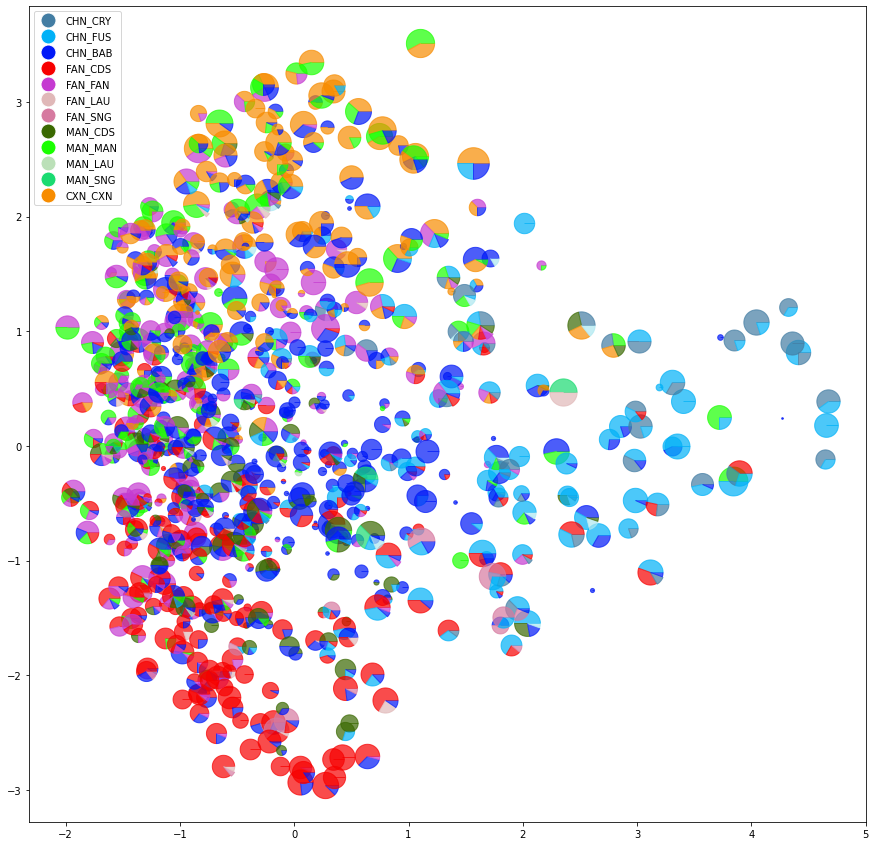}
            \caption{\small PCA, ground-truth}   
            \label{fig:mean and std of net14}
        \end{subfigure}
        \begin{subfigure}[b]{0.27\textwidth}  
          \includegraphics[width=\textwidth]{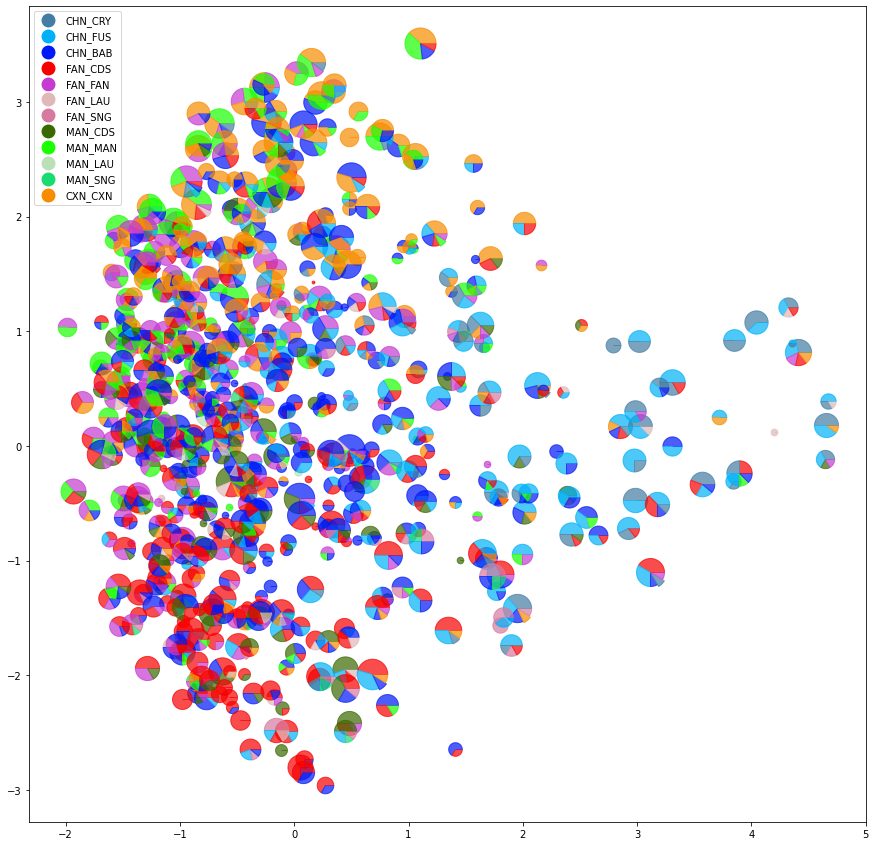}
            \caption{\small PCA, W2V2-generated}   
            \label{fig:mean and std of net24}
        \end{subfigure}
        \begin{subfigure}[b]{0.27\textwidth}   
        \centering 
         \includegraphics[width=\textwidth]{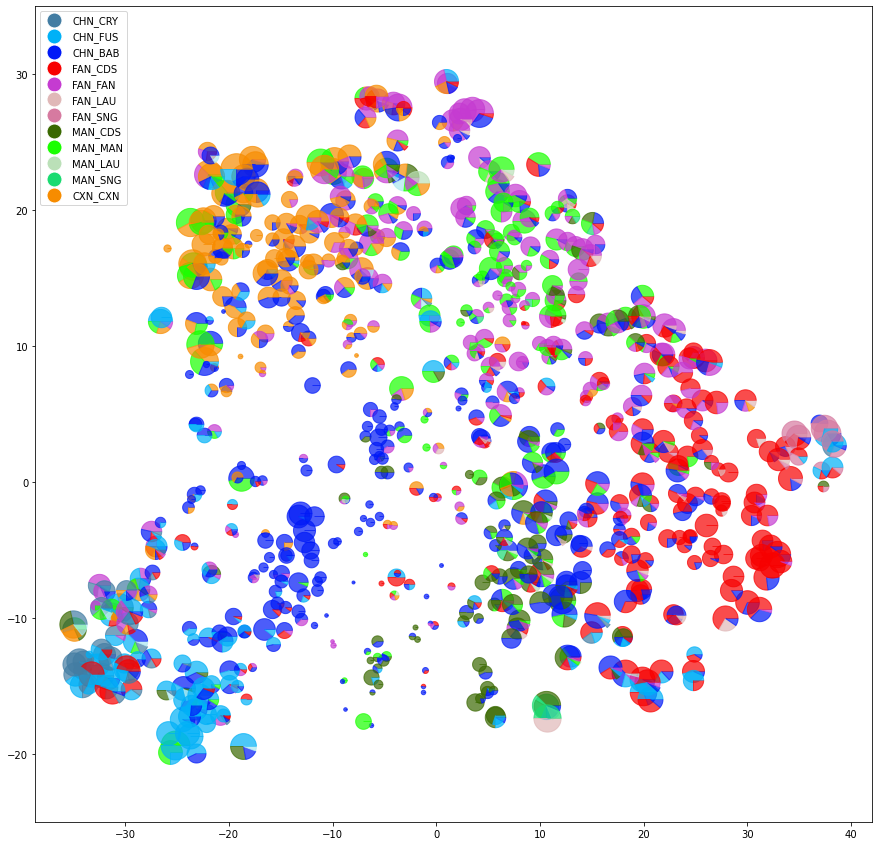}
            \caption{\small T-SNE, ground-truth}             
            \label{fig:mean and std of net34}
        \end{subfigure}
        \begin{subfigure}[b]{0.27\textwidth}   
         \includegraphics[width=\textwidth]{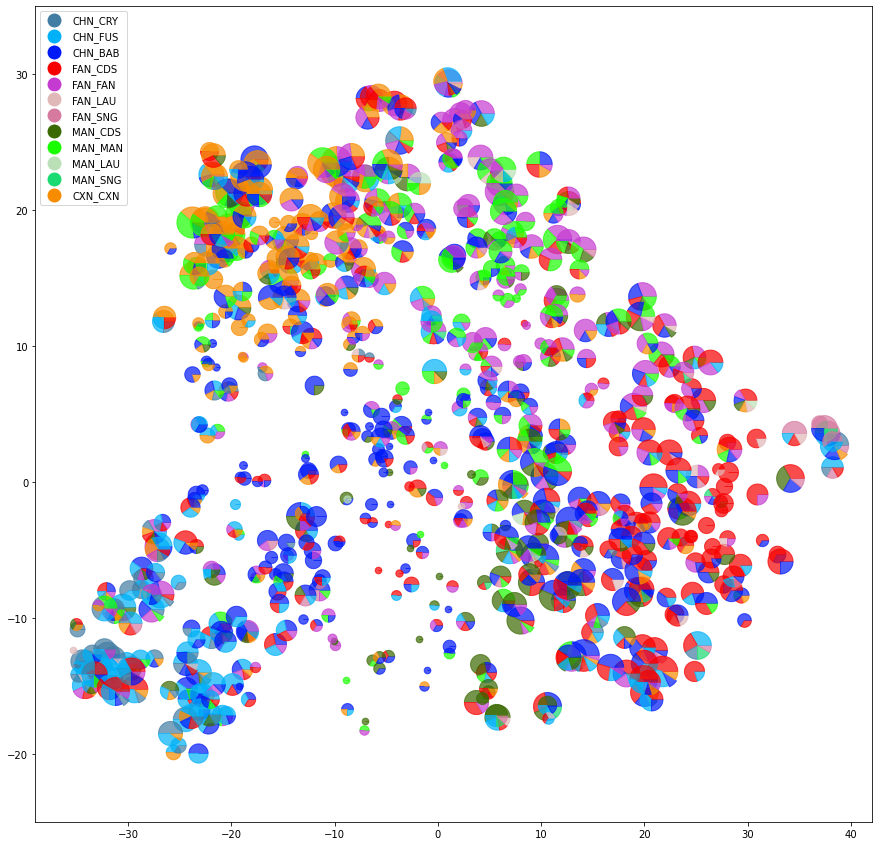}
            \caption{\small T-SNE, W2V2-generated}    
            \label{fig:mean and std of net34}
        \end{subfigure}
        \vspace{-0.2cm}
        \caption{ 
        PCA/T-SNE visualizations comparisons between ground-truth and W2V2-generated labels. Each data point is represented as a pie chart of percentage of different vocalization types over a 30s-long sequence of W2V2 features. Size of data point indicates total vocalization time fraction over each 30s interval; larger size of data point indicates longer vocalization period, and vice versa. 
        Colors used to represent each vocalization class are the following: (1) CHN (blue): 
        \texttt{CRY}
        \begin{tikzpicture}
        \fill[fill=CHN_CRY](0,0) circle (0.15); 
        \end{tikzpicture},
        \texttt{FUS}
        \begin{tikzpicture}
        \fill[fill=CHN_FUS](0,0) circle (0.15); 
        \end{tikzpicture},
        \texttt{BAB}
        \begin{tikzpicture}
        \fill[fill=CHN_BAB](0,0) circle (0.15); 
        \end{tikzpicture};
        (2) FAN (red): \texttt{CDS}
        \begin{tikzpicture}
        \fill[fill=FAN_CDS](0,0) circle (0.15); 
        \end{tikzpicture},
        \texttt{FAN} (\texttt{ADS} for FAN tier)
        \begin{tikzpicture}
        \fill[fill=FAN_FAN](0,0) circle (0.15); 
        \end{tikzpicture},
        \texttt{LAU}
        \begin{tikzpicture},
        \fill[fill=FAN_LAU](0,0) circle (0.15); 
        \end{tikzpicture},
        \texttt{SNG}
        \begin{tikzpicture}
        \fill[fill=FAN_SNG](0,0) circle (0.15); 
        \end{tikzpicture};
        (3) MAN (green): 
        \texttt{CDS}
        \begin{tikzpicture}
        \fill[fill=MAN_CDS](0,0) circle (0.15); 
        \end{tikzpicture},
        \texttt{MAN} (\texttt{ADS} for MAN tier)
        \begin{tikzpicture}
        \fill[fill=MAN_MAN](0,0) circle (0.15); 
        \end{tikzpicture},
        \texttt{LAU}
        \begin{tikzpicture}
        \fill[fill=MAN_LAU](0,0) circle (0.15); 
        \end{tikzpicture},
        \texttt{SNG}
        \begin{tikzpicture}
        \fill[fill=MAN_SNG](0,0) circle (0.15); 
        \end{tikzpicture};
        (4) CXN (orange): \texttt{CXN}
        \begin{tikzpicture}
        \fill[fill=CXN_CXN](0,0) circle (0.15); 
        \end{tikzpicture}.             
        } 
        \label{fig:ground_truth_w2v2}
    \end{figure}
\begin{figure*}[t]
        \centering
        \begin{subfigure}[b]{0.26\textwidth}   
         \includegraphics[width=\textwidth]{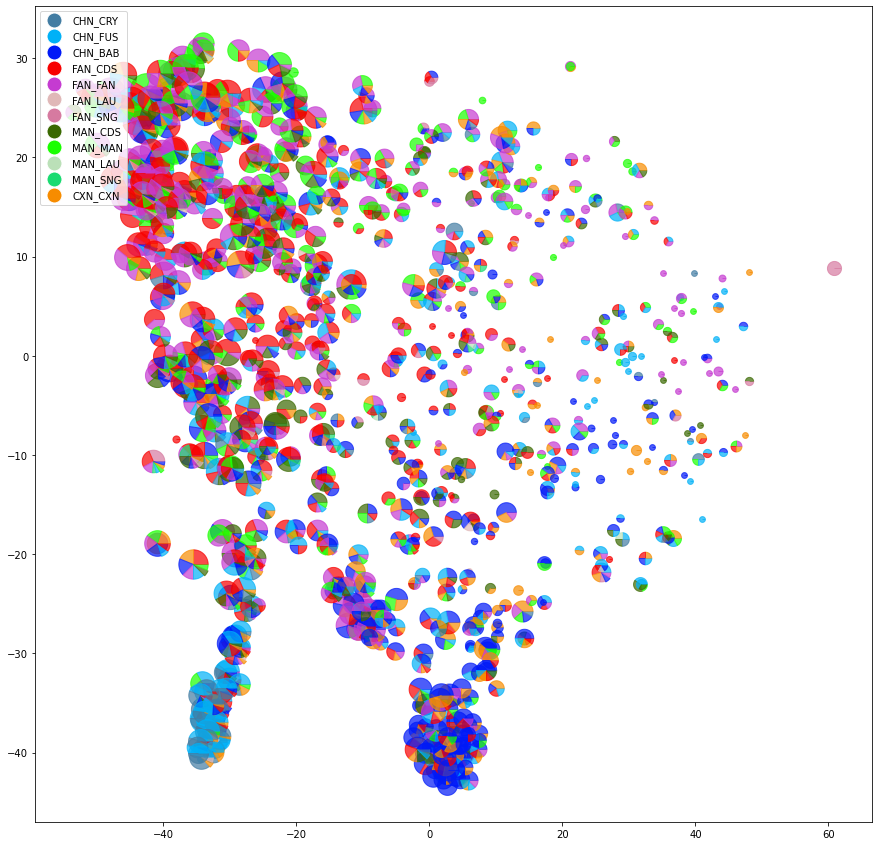}
            \caption{3.3-month} 
            \label{fig:3.3-month}
        \end{subfigure}        
        \begin{subfigure}[b]{0.26\textwidth}
          \includegraphics[width=\textwidth]{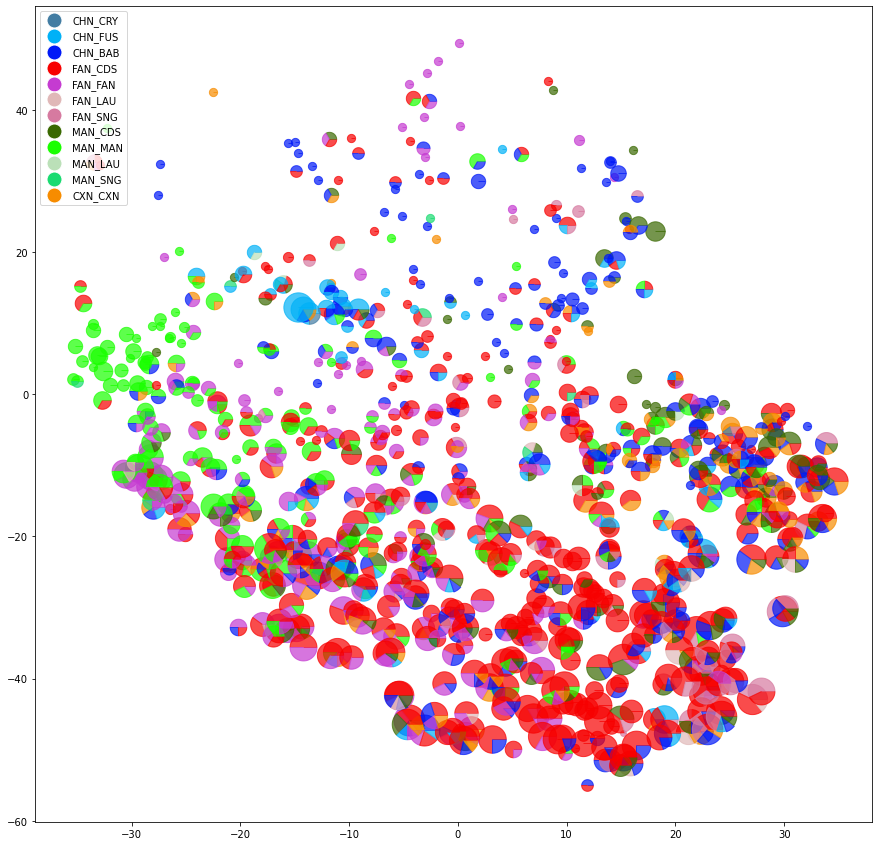}
            \caption{3.8-month}  
            \label{fig:3.8-month}
        \end{subfigure}
        \begin{subfigure}[b]{0.26\textwidth}  
           \includegraphics[width=\textwidth]{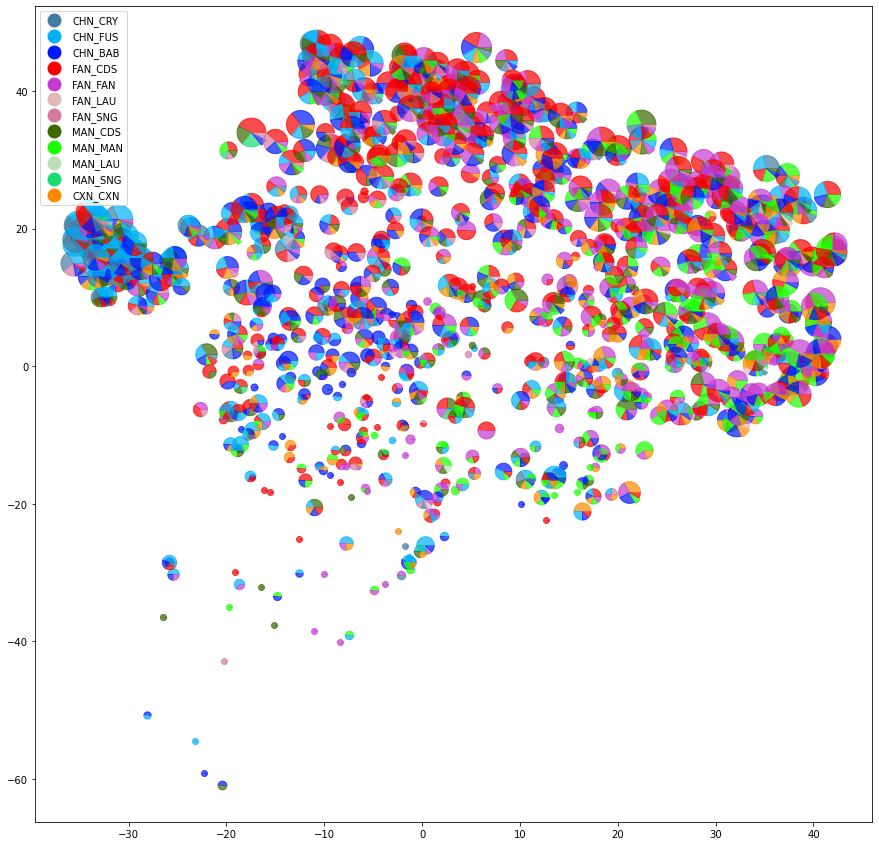}
            \caption{5.8-month}  
            \label{fig:5.8-month}
        \end{subfigure}
        \begin{subfigure}[b]{0.26\textwidth}  
           \includegraphics[width=\textwidth]{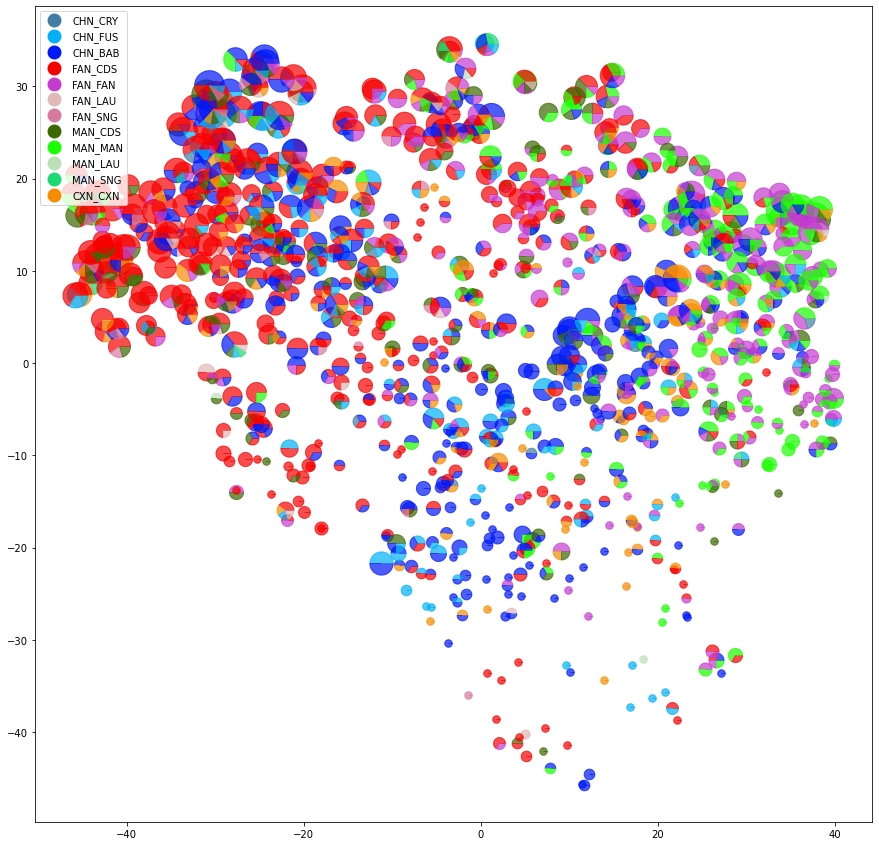}
            \caption{9.5-month} 
            \label{fig:9.5-month}
        \end{subfigure}
        \begin{subfigure}[b]{0.26\textwidth}   
         \includegraphics[width=\textwidth]{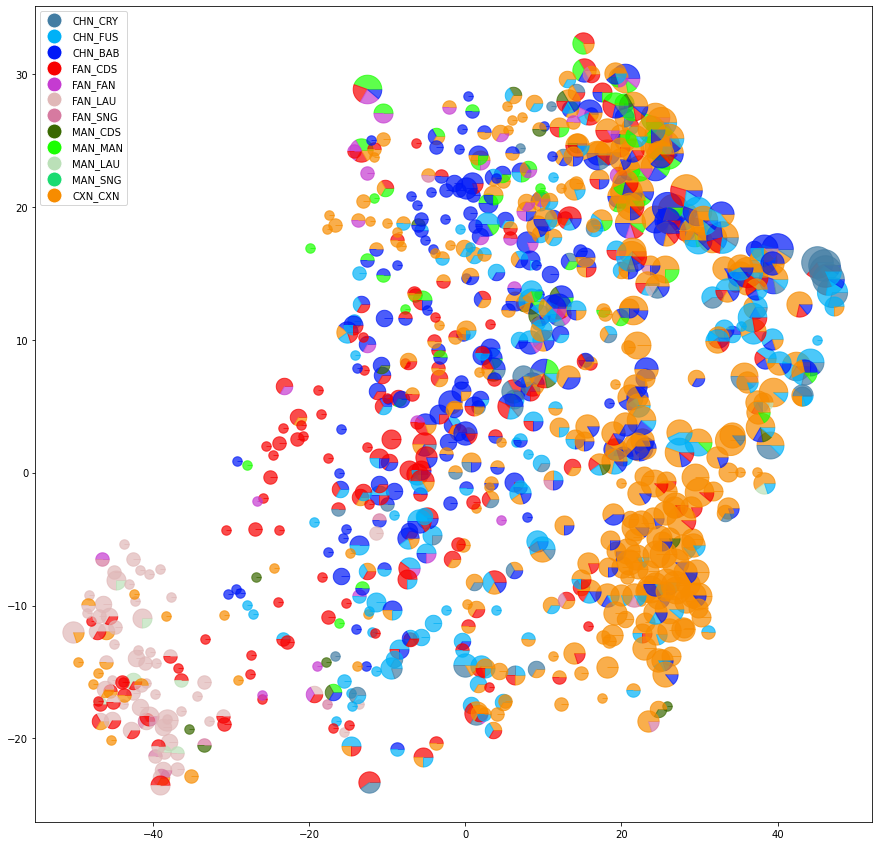}
            \caption{10.0-month}   
            \label{fig:10.0-month}
        \end{subfigure}
        \begin{subfigure}[b]{0.26\textwidth}   
         \includegraphics[width=\textwidth]{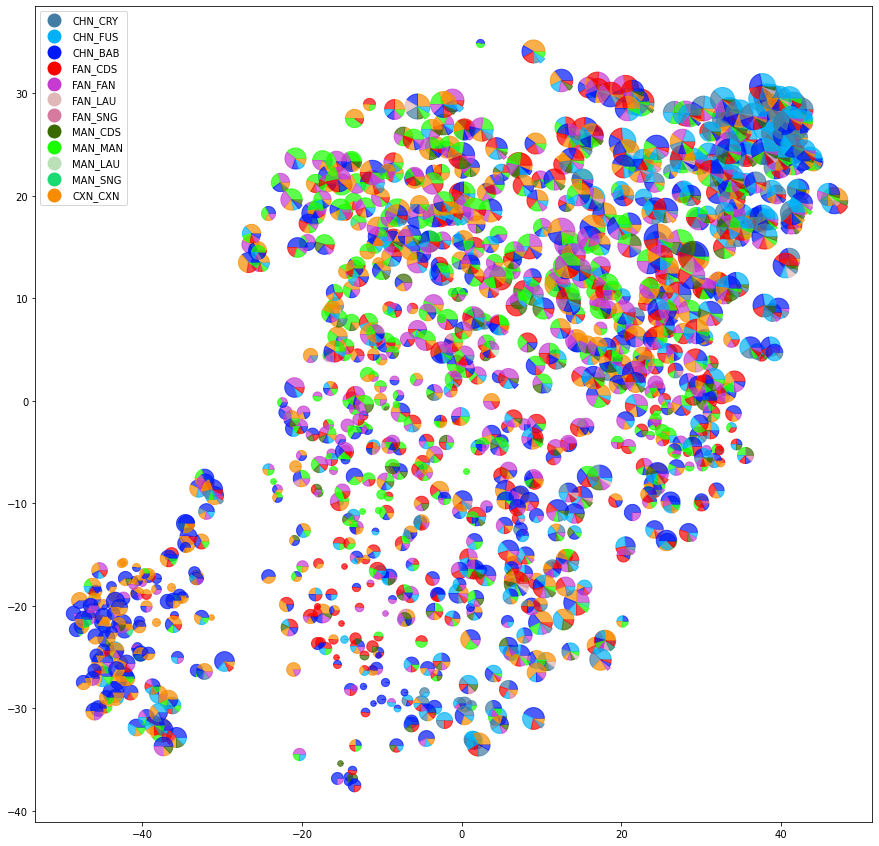}
            \caption{14.0-month}   
            \label{fig:14.0-month}
        \end{subfigure}       
        \vspace{-0.2cm}
        \caption{T-SNE visualizations of family-infant vocalization interaction patterns over 6 different families. Age of infant is indicated under each visualization. }
        \label{fig:unlabeld_audio}
        \vspace{-0.6cm}
        \end{figure*}

\vspace{-0.2cm}
\subsection{Comparisons of visualizations between using ground-truth labels and W2V2-generated labels}
\label{sec:comparisons_ground_truth_w2v2}
For all labeled audio, we compare visualizations using both ground-truth human-annotated labels and W2V2-generated labels. W2V2 is used to generate labels for intervals of 2 seconds starting every 0.2 second. To maximize accuracy of W2V2-generated labels, we label an interval as silence if (1) maximum softmax probability is below 0.8 for either the SD tier or corresponding vocalization classification tier, or (2) energy is below the threshold calculated in data prepossessing step (see section~\ref{sec:data}). 
If overlapped intervals are labeled differently, the middle time point is used to redefine the boundaries of conflicted intervals (e.g., if intervals 0-2s and 0.4-2.4s are labeled as \texttt{BAB} and \texttt{CRY} respectively, then the boundaries of labeled intervals are redefined to 0-1.2s (\texttt{BAB}) and 1.2-2.4s (\texttt{CRY}) respectively to avoid conflict labeling). 
Figure \ref{fig:ground_truth_w2v2} shows the comparisons of PCA/T-SNE visualizations between ground-truth and W2V2-generated labels. 
For ground-truth labels, both PCA and T-SNE show clear patterns among different types of infant/parent vocalization interaction patterns, including 
(1) FAN/MAN talks to CHN in \texttt{CDS} while CHN babbles (\texttt{BAB}) to respond, or CHN babbles (\texttt{BAB}) without engaging with FAN/MAN's \texttt{CDS},
(2) FAN/MAN communicates with each other in \texttt{ADS}, (3) CXN and MAN communicate with CHN together, (4) CHN is continuously babbling (\texttt{BAB}), expressing positive emotion, (5) CHN is continuously crying (\texttt{CRY})/fussing (\texttt{FUS}), expressing negative emotion, and (6) FAN occasionally comforts CHN with singing (\texttt{SNG}) or \texttt{CDS} if CHN is crying (\texttt{CRY}) or fussing (\texttt{FUS}). 
Despite that W2V2-generated labels produce slightly more noisy pie chart-based data points (e.g., samples of \texttt{BAB} are incorrectly generated) than ground-truth labels, W2V2-generated labels also show similar and clear patterns of family-infant vocalization interaction types. 
Compared with PCA, T-SNE has advantages of (1) showing more clear regions of small size data points with less active vocalization intervals and (2) producing more distinct boundaries among different interaction patterns, such as MAN talks to CHN in \texttt{CDS}. 
\vspace{-0.3cm}
\subsection{Visualizations over unlabeled audio of each family}
\vspace{-0.1cm}
Because Figure \ref{fig:ground_truth_w2v2} qualitatively verifies the correctness of W2V2-generated labels, therefore it is meaningful  to 
apply the same BoAW approach using pretrained codebooks on all collected unlabeled audio over each individual family. To avoid overcrowded data points in visualizations, we use k-means to cluster the BoAW histogram vectors from each family into 8 clusters and randomly select 100 data points from each cluster to present. Figure \ref{fig:unlabeld_audio} shows T-SNE visualizations over 6 different families.  We observe relatively clean boundaries of interaction patterns within each family, and these patterns differ considerably across families, illustrating the success of our BoAW algorithms for creating visualizations in a semi-supervised learning setting.
All families are characterized using BoAW histograms computed over the same set of audio words,
so each family pattern can be interpreted as a subset of the multi-family patterns
shown in Figure \ref{fig:ground_truth_w2v2}.  
In Figure \ref{fig:unlabeld_audio}, visualizations show significantly different interaction patterns among different families in spite of similar ages of infants. 
For example, a 3.3-month-old infant family pattern (Figure \ref{fig:3.3-month}) shows that the infant (CHN) is actively vocalizing \texttt{BAB}/\texttt{CRY}/\texttt{FUS}, and parents(FAN/MAN) communicate
alternatively between \texttt{CDS} and \texttt{ADS}, while a 3.8-month-old infant family pattern (Figure \ref{fig:3.8-month}) indicates that the infant (CHN) exhibits fewer vocalizations, and the majority family interaction type is that adults (mostly FAN, occasionally MAN) talk
in \texttt{CDS} or \texttt{ADS}. Likewise, noticeable differences emerge in family-infant interaction patterns for a 9.5-month-old (Figure \ref{fig:9.5-month}) versus is a 10.0-month-old (Figure \ref{fig:10.0-month}). In Figure \ref{fig:9.5-month}, we observe that (1) FAN/CXN communicates with CHN in \texttt{CDS} while CHN vocalizes (\texttt{BAB}/\texttt{FUS}), (2) CHN vocalizes to himself/herself without communicating to other family members, and (3) FAN/MAN talk with each other in \texttt{ADS}. In Figure \ref{fig:10.0-month}, we observe that CXN is very actively vocalizing, and FAN occasionally sings (\texttt{SNG}) to the CHN, and the CHN often babbles (\texttt{BAB}) or cries (\texttt{CRY}) herself/himself. Compared with younger infants, a 14.0-month-old infant (Figure \ref{fig:14.0-month}) shows complex interaction patterns with family members, as many pie chart data points mix multiple vocalization types from different family members. 
\vspace{-0.3cm}
\section{Future Work \& Conclusion}
\vspace{-0.2cm}
In this study, we demonstrate an innovative way to condense and visualize rich information of complex family-infant vocalization interaction patterns in a 2-dimensional space based on W2V2 features using BoAW approach. We qualitatively evaluate and approve the visualization quality on labeled family audio of W2V2-generated labels by comparing against ground-truth labels. We apply the same workflow to generate visualizations of unlabeled audio over each individual family and obtain useful high-level audio understanding for different family-infant interaction patterns. 

In the future, we aim to advance further the meaningfulness of family interaction visualizations by possibly incorporating other useful features to distinguish between coordinated parent-infant interactions (e.g., displays of positive affect and turn-taking between the infant and parent) as well as uncoordinated interactions (e.g., parent is unresponsive to infant’s cues; parent and infant interrupt or talk over each other).

\bibliographystyle{IEEEtran}

\bibliography{mybib}

\begin{thebibliography}{10}
\providecommand{\url}[1]{#1}
\csname url@samestyle\endcsname
\providecommand{\newblock}{\relax}
\providecommand{\bibinfo}[2]{#2}
\providecommand{\BIBentrySTDinterwordspacing}{\spaceskip=0pt\relax}
\providecommand{\BIBentryALTinterwordstretchfactor}{4}
\providecommand{\BIBentryALTinterwordspacing}{\spaceskip=\fontdimen2\font plus
\BIBentryALTinterwordstretchfactor\fontdimen3\font minus
  \fontdimen4\font\relax}
\providecommand{\BIBforeignlanguage}[2]{{%
\expandafter\ifx\csname l@#1\endcsname\relax
\typeout{** WARNING: IEEEtran.bst: No hyphenation pattern has been}%
\typeout{** loaded for the language `#1'. Using the pattern for}%
\typeout{** the default language instead.}%
\else
\language=\csname l@#1\endcsname
\fi
#2}}
\providecommand{\BIBdecl}{\relax}
\BIBdecl

\bibitem{bitsko2016health}
R.~H. Bitsko, J.~R. Holbrook, L.~R. Robinson, J.~W. Kaminski, R.~Ghandour,
  C.~Smith, and G.~Peacock, ``Health care, family, and community factors
  associated with mental, behavioral, and developmental disorders in early
  childhood—united states, 2011--2012,'' \emph{Morbidity and Mortality Weekly
  Report}, vol.~65, no.~9, pp. 221--226, 2016.

\bibitem{coercive}
\BIBentryALTinterwordspacing
G.~R. Patterson, ``The early development of coercive family process,'' in
  \emph{J. B. Reid, G. R. Patterson, \& J. Snyder (Eds.), Antisocial behavior
  in children and adolescents: A developmental analysis and model for
  intervention, American Psychological Association.}, 2002, pp. 25--44.
  [Online]. Available: \url{https://doi.org/10.1037/10468-002}
\BIBentrySTDinterwordspacing

\bibitem{dunn1983sibling}
J.~Dunn, ``Sibling relationships in early childhood,'' \emph{Child
  development}, pp. 787--811, 1983.

\bibitem{kolak2011sibling}
A.~M. Kolak and B.~L. Volling, ``Sibling jealousy in early childhood:
  Longitudinal links to sibling relationship quality,'' \emph{Infant and Child
  Development}, vol.~20, no.~2, pp. 213--226, 2011.

\bibitem{volling2002emotion}
B.~L. Volling, N.~L. McElwain, and A.~L. Miller, ``Emotion regulation in
  context: The jealousy complex between young siblings and its relations with
  child and family characteristics,'' \emph{Child development}, vol.~73, no.~2,
  pp. 581--600, 2002.

\bibitem{dadds1992childhood}
M.~R. Dadds, M.~R. Sanders, M.~Morrison, and M.~Rebgetz, ``Childhood depression
  and conduct disorder: Ii. an analysis of family interaction patterns in the
  home.'' \emph{Journal of Abnormal Psychology}, vol. 101, no.~3, p. 505, 1992.

\bibitem{ward2016parent}
M.~A. Ward, J.~Theule, and K.~Cheung, ``Parent--child interaction therapy for
  child disruptive behaviour disorders: A meta-analysis,'' in \emph{Child \&
  Youth Care Forum}, vol.~45, no.~5.\hskip 1em plus 0.5em minus 0.4em\relax
  Springer, 2016, pp. 675--690.

\bibitem{LENA}
X.~D., J.~A. Richards, and J.~Gilkerson, ``Automated analysis of child phonetic
  production using naturalistic recordings.'' \emph{Journal of Speech, Language
  \& Hearing Research}, vol.~57, no.~5, pp. 1638 -- 1650, 2014.

\bibitem{anders2020automatic}
F.~Anders, M.~Hlawitschka, and M.~Fuchs, ``Automatic classification of infant
  vocalization sequences with convolutional neural networks,'' \emph{Speech
  Communication}, vol. 119, pp. 36--45, 2020.

\bibitem{anders2020comparison}
------, ``Comparison of artificial neural network types for infant vocalization
  classification,'' \emph{IEEE/ACM Transactions on Audio, Speech, and Language
  Processing}, vol.~29, pp. 54--67, 2020.

\bibitem{ji2021review}
C.~Ji, T.~B. Mudiyanselage, Y.~Gao, and Y.~Pan, ``A review of infant cry
  analysis and classification,'' \emph{EURASIP Journal on Audio, Speech, and
  Music Processing}, vol. 2021, no.~1, pp. 1--17, 2021.

\bibitem{li2021analysis}
J.~Li, M.~Hasegawa-Johnson, and N.~L. McElwain, ``Analysis of acoustic and
  voice quality features for the classification of infant and mother
  vocalizations,'' \emph{Speech Communication}, vol. 133, pp. 41--61, 2021.

\bibitem{gujral2019leveraging}
A.~Gujral, K.~Feng, G.~Mandhyan, N.~Snehil, and T.~Chaspari, ``Leveraging
  transfer learning techniques for classifying infant vocalizations,'' in
  \emph{2019 IEEE EMBS International Conference on Biomedical \& Health
  Informatics (BHI)}.\hskip 1em plus 0.5em minus 0.4em\relax IEEE, 2019, pp.
  1--4.

\bibitem{wav2vec}
A.~Baevski, H.~Zhou, A.~Mohamed, and M.~Auli, ``wav2vec 2.0: A framework for
  self-supervised learning of speech representations,'' \emph{arXiv e-prints},
  2020.

\bibitem{li2021accent}
J.~Li, V.~Manohar, P.~Chitkara, A.~Tjandra, M.~Picheny, F.~Zhang, X.~Zhang, and
  Y.~Saraf, ``Accent-robust automatic speech recognition using supervised and
  unsupervised wav2vec embeddings,'' \emph{arXiv preprint arXiv:2110.03520},
  2021.

\bibitem{schneider2019wav2vec}
S.~Schneider, A.~Baevski, R.~Collobert, and M.~Auli, ``wav2vec: Unsupervised
  pre-training for speech recognition,'' \emph{arXiv preprint
  arXiv:1904.05862}, 2019.

\bibitem{pepino2021emotion}
L.~Pepino, P.~Riera, and L.~Ferrer, ``Emotion recognition from speech using
  wav2vec 2.0 embeddings,'' \emph{arXiv preprint arXiv:2104.03502}, 2021.

\bibitem{yuan2021role}
J.~Yuan, X.~Cai, R.~Zheng, L.~Huang, and K.~Church, ``The role of phonetic
  units in speech emotion recognition,'' \emph{arXiv preprint
  arXiv:2108.01132}, 2021.

\bibitem{wang2021fine}
Y.~Wang, A.~Boumadane, and A.~Heba, ``A fine-tuned wav2vec 2.0/hubert benchmark
  for speech emotion recognition, speaker verification and spoken language
  understanding,'' \emph{arXiv preprint arXiv:2111.02735}, 2021.

\bibitem{bulgarelli2020look}
F.~Bulgarelli and E.~Bergelson, ``Look who’s talking: A comparison of
  automated and human-generated speaker tags in naturalistic day-long
  recordings,'' \emph{Behavior Research Methods}, vol.~52, no.~2, pp. 641--653,
  2020.

\bibitem{boersma2006praat}
P.~Boersma, ``Praat: doing phonetics by computer,'' \emph{http://www. praat.
  org/}, 2006.

\bibitem{speechbrain}
M.~Ravanelli, T.~Parcollet, P.~Plantinga, A.~Rouhe, S.~Cornell, L.~Lugosch,
  C.~Subakan, N.~Dawalatabad, A.~Heba, J.~Zhong, J.-C. Chou, S.-L. Yeh, S.-W.
  Fu, C.-F. Liao, E.~Rastorgueva, F.~Grondin, W.~Aris, H.~Na, Y.~Gao, R.~D.
  Mori, and Y.~Bengio, ``{SpeechBrain}: A general-purpose speech toolkit,''
  2021, arXiv:2106.04624.

\bibitem{zhang2010understanding}
Y.~Zhang, R.~Jin, and Z.-H. Zhou, ``Understanding bag-of-words model: a
  statistical framework,'' \emph{International Journal of Machine Learning and
  Cybernetics}, vol.~1, no.~1, pp. 43--52, 2010.

\bibitem{zhang2018unsupervised}
L.~Zhang, J.~Han, and S.~Deng, ``Unsupervised temporal feature learning based
  on sparse coding embedded boaw for acoustic event recognition.'' in
  \emph{INTERSPEECH}, 2018, pp. 3284--3288.

\bibitem{pancoast2012bag}
S.~Pancoast and M.~Akbacak, ``Bag-of-audio-words approach for multimedia event
  classification,'' SRI International Menlo Park United States, Tech. Rep.,
  2012.

\bibitem{schmitt2016border}
M.~Schmitt, F.~Ringeval, and B.~W. Schuller, ``At the border of acoustics and
  linguistics: Bag-of-audio-words for the recognition of emotions in speech.''
  in \emph{Interspeech}, 2016, pp. 495--499.

\bibitem{han2018bags}
J.~Han, Z.~Zhang, M.~Schmitt, Z.~Ren, F.~Ringeval, and B.~Schuller, ``Bags in
  bag: Generating context-aware bags for tracking emotions from speech,'' in
  \emph{Interspeech 2018}.\hskip 1em plus 0.5em minus 0.4em\relax ISCA, 2018,
  pp. 3082--3086.

\bibitem{schuller2019interspeech}
B.~Schuller, A.~Batliner, C.~Bergler, F.~B. Pokorny, J.~Krajewski, M.~Cychosz,
  R.~Vollmann, S.-D. Roelen, S.~Schnieder, E.~Bergelson \emph{et~al.}, ``The
  interspeech 2019 computational paralinguistics challenge: Styrian dialects,
  continuous sleepiness, baby sounds \& orca activity,'' 2019.

\bibitem{schuller2018interspeech}
B.~Schuller, S.~Steidl, A.~Batliner, P.~B. Marschik, H.~Baumeister, F.~Dong,
  S.~Hantke, F.~B. Pokorny, E.-M. Rathner, K.~D. Bartl-Pokorny \emph{et~al.},
  ``The interspeech 2018 computational paralinguistics challenge: atypical and
  self-assessed affect, crying and heart beats,'' 2018.

\bibitem{schuller2017interspeech}
B.~Schuller, S.~Steidl, A.~Batliner, E.~Bergelson, J.~Krajewski, C.~Janott,
  A.~Amatuni, M.~Casillas, A.~Seidl, M.~Soderstrom \emph{et~al.}, ``The
  interspeech 2017 computational paralinguistics challenge: Addressee, cold \&
  snoring,'' in \emph{Computational Paralinguistics Challenge (ComParE),
  Interspeech 2017}, 2017, pp. 3442--3446.

\bibitem{schuller2021interspeech}
B.~W. Schuller, A.~Batliner, C.~Bergler, C.~Mascolo, J.~Han, I.~Lefter,
  H.~Kaya, S.~Amiriparian, A.~Baird, L.~Stappen \emph{et~al.}, ``The
  interspeech 2021 computational paralinguistics challenge: Covid-19 cough,
  covid-19 speech, escalation \& primates,'' \emph{arXiv preprint
  arXiv:2102.13468}, 2021.

\bibitem{van2008visualizing}
L.~Van~der Maaten and G.~Hinton, ``Visualizing data using t-sne.''
  \emph{Journal of machine learning research}, vol.~9, no.~11, 2008.

\bibitem{schmitt2017openxbow}
M.~Schmitt and B.~Schuller, ``Openxbow: introducing the passau open-source
  crossmodal bag-of-words toolkit,'' in \emph{Proc. ICML}, 2017.

\bibitem{scikit-learn}
F.~Pedregosa, G.~Varoquaux, A.~Gramfort, V.~Michel, B.~Thirion, O.~Grisel,
  M.~Blondel, P.~Prettenhofer, R.~Weiss, V.~Dubourg, J.~Vanderplas, A.~Passos,
  D.~Cournapeau, M.~Brucher, M.~Perrot, and E.~Duchesnay, ``Scikit-learn:
  Machine learning in {P}ython,'' \emph{Journal of Machine Learning Research},
  vol.~12, pp. 2825--2830, 2011.

\end{thebibliography}


\end{document}